\documentclass[amsfonts,pre,aps]{revtex4}
\usepackage{amsmath}\usepackage{amssymb}
\begin{document}
\title{A brief critique of the Adam-Gibbs entropy model} 
\author{Jeppe C. Dyre, Tina Hechsher, and Kristine Niss}
\affiliation{DNRF Centre ``Glass and Time,'' IMFUFA (Building 27), Department of Sciences, Roskilde University, Postbox 260, DK-4000 Roskilde,
Denmark} 
\date{\today}
\newcommand{\sco}{S_{\rm c}}
\newcommand{\sv}{S_{\rm vib}}
\newcommand{\sex}{S_{\rm exc}}
\begin{abstract}
This paper critically reviews the entropy model proposed by Adam and Gibbs in 1965 for explaining the dramatic temperature dependence of glass-forming liquids' average relaxation time, one of the most influential models during the last three decades. We discuss the Adam-Gibbs model's theoretical bases as well as the reported experimental model confirmations; in the process of doing this a number of problems with the model are identified.
\end{abstract}

\pacs{64.70.Pf}
\maketitle

\section{Introduction}
The ability to form glasses is a universal property of liquids, i.e., any liquid forms a glass when supercooled rapidly enough to avoid crystallization \cite{tam25,sim31,kau48,joh74,har76,bra85,sch86,ang91,deb96,ang00,alb01,don01,das04,bin05,dyr06}. Glass formation is an example of the ``falling-out-of-equilibrium'' that takes place for any system the relaxation time of which exceeds laboratory time scales \cite{dyr08}. In our opinion, this phenomenon does not in itself present subtle scientific questions, although there may well be interesting relaxations taking place at $T_g$ affecting details of the glass structure \cite{dyr87}. The ultraviscous liquid in metastable equilibrium above $T_g$, on the other hand, does present fundamental scientific challenges. The two most important questions relating to the ultraviscous liquid phase are: 1) What causes the non-exponential relaxations usually observed? 2) What causes the non-Arrhenius temperature dependence of the average (alpha) relaxation time $\tau$? This paper addresses one of the classical answers to the latter question. 

Most viscous liquids require temperature dependence of the activation energy $\Delta E=\Delta E(T)$ if the Arrhenius expression is accepted,

\begin{equation}\label{arrh}
\tau(T)\,=\,
\tau_0\exp\left(\frac{\Delta E(T)}{k_BT}\right)\,.
\end{equation}
Molten pure silica and a few other liquids have almost temperature-independent activation energy, but for all other liquids the activation energy increases upon cooling. A measure of how fast the activation energy increases is the ``temperature index'' defined \cite{dyr04} by $I(T)=-d\ln\Delta E/d\ln T\ge 0$. The standard measure of the degree of non-Arrhenius behavior is Angell's fragility $m$ defined by $m=d\log\tau/d(T_g/T)|_{T=T_g}$ \cite{pla91,boh93,ruo04}, a quantity that only refers to liquid properties right at $T_g$. If the glass transition temperature by definition is taken as the temperature where $\tau=100{\rm s}$ and $\tau_0=10^{-14}{\rm s}$, Arrhenius behavior corresponds to $m=16$. In the index terminology Arrhenius behavior corresponds to $I=0$. Generally, the following relation allows one to calculate the fragility from the index at $T_g$: $m=16[1+I(T_g)]$ \cite{dyr04}.

There is no general agreement about the origin of the non-Arrhenius behavior of viscous liquids. It may well be that there no simple, universally valid model or theory exists, but many workers in the field including ourselves until proved otherwise prefer to assume that such a model exists. This is a reasonable assumption, because ultraviscous liquids approaching the glass transition have physical properties that do not depend on whether the liquid is bonded by covalent bonds, ionic bonds, van der Waals bonds, hydrogen bonds, or metallic bonds 
\cite{tam25,sim31,kau48,joh74,har76,bra85,sch86,ang91,deb96,ang00,alb01,don01,das04,bin05,dyr06} (we prefer to exclude the often studied polymer glass transition because it is not a liquid-glass transition, though it is noteworthy that this transition has several properties similar to those of the liquid-glass transition). 

Whenever an important scientific problem is unsolved, there is usually not one, but many models claiming to solve the problem. The non-Arrhenius behavior of glass-forming liquids is no exception. Classical phenomenological models relate the relaxation time to macroscopic liquid properties, like the configurational entropy \cite{gib58,ada65}, the free volume \cite{doo51,coh59,kov63,gre81}, the energy \cite{bra85,dyr87,gol72,nem78,bou92,dyr95,die97}, or the high-frequency elastic constants \cite{tob43,moo57,bue59,nem68,dyr96}. More recently, these models were supplemented by theories that generally have a more fundamental basis like, e.g., the mode coupling theory \cite{das04,got92}, the random-first-order-transition theory (RFOT) \cite{bou04,lub07}, energy-landscape based models 
\cite{gol69,sti95,deb01,wal03,sci05}, frustration-based approaches \cite{tar05}, the entropic barrier hopping theory \cite{sch04}, kinetically constrained models \cite{fre88,gar03}, etc. 

This paper addresses one of the most popular classical models, the Adam-Gibbs entropy model \cite{ada65}. We first briefly review the model and how it was traditionally supported by experiment (Sec. II). In Sec. III critiques of the model are presented, relating to the model's theoretical basis as well as its experimental validation. Many of these critiques have been made before, but we felt it would be useful to collect them into one paper. Section IV concludes.

\section{The Adam-Gibbs entropy model}

\subsection{Assumptions and model prediction}

According to the Adam-Gibbs model the liquid's relaxation time is controlled by its configurational entropy $\sco(T)$. This quantity is defined by subtracting the vibrational entropy, $\sv(T)$, from the entropy $S$: $\sco(T)=S(T)-\sv(T)$. This separation of entropy into two contributions is much in the spirit of the energy landscape paradigm subsequently formulated by Goldstein \cite{gol69} and Stillinger \cite{sti83}, where vibrations around a potential energy minimum (an inherent state) are occasionally interrupted by thermally activated transitions to another minimum.

The Adam-Gibbs model's activation energy is characterized by the property

\begin{equation}\label{entropy_model}
\Delta E(T)
\,\propto\,\frac{1}{\sco(T)}\,.
\end{equation}
This is justified as follows. Any molecular rearrangement is a thermally activated transition that involves all molecules of a ``cooperatively rearranging region.'' Such a region is defined as a ``subsystem of the sample which, upon a sufficient fluctuation in energy (or, more correctly, enthalpy), can rearrange into another configuration independently of its environment.'' Three crucial ideas/assumptions go into the model: 
\begin{enumerate}
\item The activation energy is proportional to region volume. This is justified by writing the change in Gibbs free energy upon activation as a chemical potential change $\Delta\mu$ times volume and assuming that ``in a good approximation the dependence of $\Delta\mu$ on temperature and region volume can be neglected.'' 
\item There is a lower limit to the size of a cooperatively rearranging region since it must have at least two configurations ``available to it, one in which the region resides before the transition and another one to which it may move.'' 
\item The cooperatively rearranging regions are ``independent and equivalent subsystems,'' i.e., there are only insignificant interactions of any given region with its surroundings.
\end{enumerate}

\subsection{The model's attractive scenario}

The Adam-Gibbs model connects two of the most fundamental and intriguing concepts of physics: {\it Entropy} and {\it Time}. The model is aesthetically most attractive by having this property; the only other {\it quantitative} connection of entropy and time that we can think of is that of black hole thermodynamics as theorized by Hawking and others (the fact that entropy cannot decrease for an isolated system is a {\it qualitative} entropy-time connection). The entropy model has the further beauty of connecting the observed dramatic slowing down to the Kauzmann paradox and the theory of phase transitions. Recall that the Kauzmann paradox is the observation that the supercooled liquid's excess entropy $\sex$ (the liquid entropy minus the crystal entropy at the same temperature) extrapolates to zero at a temperature $T_K$ not far below $T_g$ \cite{kau48}. Unless something rather dramatic happens invalidating this extrapolation, the liquid's entropy would fall below the crystal's if the liquid could be equilibrated close to and below $T_K$. But if -- as often done in practice -- the excess entropy is identified with the configurational entropy (a point returned to below),

\begin{equation}\label{sexeqscon}
\sex(T)\,\cong\,\sco(T)\,,
\end{equation}
the Adam-Gibbs (AG) model resolves the Kauzmann paradox: By Eq. (\ref{entropy_model}) the relaxation time diverges to infinity as the liquid is cooled towards $T_K$. This means that the liquid cannot equilibrate close to $T_K$, implying that the glass transition must take place above $T_K$ no matter how slowly the liquid is cooled.

Based on Eq. (\ref{sexeqscon}) the model presents a scenario that predicts an underlying phase transition to a state of zero configurational entropy and infinite relaxation time. Thus the model explains the dramatic relaxation-time increase as a consequence of the approach to a phase transition. The predicted slowing down extends over a broader temperature range and is much more dramatic than the usual critical slowing down for second order phase transitions where $\tau\propto |T-T_c|^{-x}$ \cite{gol92}, but the idea is the same. In this way, the paradigm of second order phase transitions comes into play for the glass transition problem.

The above explains the AG model's attraction in general, theoretical terms. The AG model's main attraction, however, is probably the fact that it appears to explain numerous experiments. We shall not detail the evidence for this here, but refer the reader to several excellent reviews \cite{ang91,ang00,alb01,entropy_vol}. In many cases the experimental evidence for the AG model relates it to the Vogel-Fulcher-Tammann (VFT) empirical equation for the relaxation time:

\begin{equation}\label{vft}
\tau(T)\,=\,
\tau_0\exp\left(\frac{A}{T-T_0}\right)\,.
\end{equation}
Close to $T_K$ the configurational entropy $\sco(T)$ may be expanded to first order: $\sco(T)\propto T-T_K$, implying that the AG model predicts

\begin{equation}\label{tkeqto}
T_K\,=\,T_0\,.
\end{equation}
This prediction has been compared to experiment on many liquids. The general picture is that the AG model is obeyed for most, if not all systems studied \cite{entropy_vol,ric84}. These include chemically quite different systems with widely differing glass transition temperatures.

\section{Critiques of the entropy model}

\subsection{Model assumptions}

As mentioned, the three basic assumptions of the AG model are: 1) The activation energy is proportional to region volume, $\Delta E(T)\propto V_{\rm reg}(T)$; 2) A region must have at least two configurations, i.e., it must have a configurational entropy at least of order $k_B$; 3) The ``region assumption'' that regions are {\it independent} and {\it equivalent} subsystems of the liquid. None of points 1)-3) are compelling: Molecular rearrangements take place in almost perfect crystals via diffusing vacancies or interstitials, and in a plastic crystal, for instance, one might well have molecular reorientations happening without either assumption 1) or 2) being obeyed. This is also an example where assumption 3) breaks down. Even if assumption 3) holds, however, it is not necessary that a region must have a minimum configurational entropy in order to allow transitions; also for a low configurational entropy a region would have many states ``available to it'' if differing energies are allowed for. 

Assumption 1), which is responsible for the non-Arrhenius behavior and the eventual relaxation time divergence as $T\rightarrow T_K$, was justified by the assumption that the chemical potential difference between initial state and the transition state (barrier) is region-size independent. The question is, however, how well defined a chemical potential difference is for this situation (particularly in view of the small region sizes inferred from experiment that makes it difficult to justify ignoring the interactions with the surroundings, see below).

Finally, returning to the region assumption 3) we note that it can only be justified if regions are very large. As an analogue, note that even for rather large ``regions,'' nucleation theory must take into account the interactions with the surrounding liquid in order to arrive at realistic predictions. It is not clear why the same should not be done in the Adam-Gibbs theory; indeed, this is done in the more sophisticated RFOT entropy model of Wolynes and coworkers \cite{bou04,lub07}.

Suppose that we nevertheless accept assumptions 1), 2) and 3) and go ahead comparing to experiment. When this is done, one typically arrives at regions containing 4-8 molecules \cite{alb01,yam98} close to the glass transition! At higher temperatures regions must be even smaller, because it is the increasing regions size upon cooling that is responsible for the non-Arrhenius behavior. The small region sizes of experiment present a serious challenge to the AG entropy model, because such small regions cannot reasonably be regarded as independent with region-region interactions that may be ignored; every molecule must interact with molecules of other regions as much as with the molecules within a given region.

Suppose that one nevertheless accepts the AG idea that the configurational entropy controls the relaxation time's temperature dependence and also accepts Eq. (\ref{sexeqscon}) that allows for the entropy model to be compared to experiment. Then, as mentioned, the relaxation time becomes infinite at $T_K$ where the equilibrium state of the liquid has zero configurational entropy --  the ``ideal glass'' state is approached \cite{ang68,gre07}. This state cannot be reached experimentally, but one may still ask what is its nature. A state of zero entropy is unique like a perfect crystal, so one would expect that some simple description of it could be given. Except for the random close packing of hard spheres -- the uniqueness of which is questioned -- we are not aware of attempts to describe the ideal glassy state in structural terms. This does not rule out that such a description exists, but one would imagine it to be fairly simple (like a quasi-crystal) and thus has been identified long ago.

\subsection{The entropy model's experimental verification}

Despite the above theoretical arguments, suppose that we AG model's prediction Eq. (\ref{entropy_model}). Unfortunately, configurational entropy cannot be measured. For many years this challenge was approached by arguing as follows: {\it The vibrational properties of glass and crystal are very similar, and very similar to the liquid's high-frequency vibrational properties (i.e., on time scales much shorter than those of the alpha (main) relaxation time). Since the crystalline state has practically zero configurational entropy, the crystal entropy provides a good estimate of the liquid's vibrational entropy. Thus by subtracting crystal entropy from liquid entropy one finds the liquid's configurational entropy (Eq. (\ref{sexeqscon})).} 

There is now a growing recognition that the above argument is problematic \cite{dyr06,gol76,mar01,joh02}. Dating back to the 1950's, in fact, it was known from sound velocity measurements that the liquid's high-frequency sound velocity is generally much more temperature dependent than that of the crystal or glass phases \cite{har76,lit59,bar72}. It is easy to understand why this is so if one adopts the simple-minded assumption that the high-frequency sound velocity is a function of density: The thermal expansion coefficient is generally considerably larger in the liquid than in the solid phases (crystal or glass). In this approach, the vibrational entropy is a (logarithmic) function of the vibrational force constants that determine the high-frequency sound velocity, so the vibrational entropy is considerably more temperature dependent in the liquid than in the crystal. This severely weakens Eq. (\ref{sexeqscon}). An illustration of the problem with Eq. (\ref{sexeqscon}) is the fact that it is not generally true that a liquid must have larger entropy than the same temperature crystal: Both in the cases involving so-called inverse melting \cite{mor92,sti03} as well as for the classical hard sphere system, the crystalline phase has larger entropy than the liquid at same thermodynamic conditions.

Suppose that we nevertheless accept Eq. (\ref{sexeqscon}). Then at the Kauzmann temperature $T_K$ there is a second order phase transition to the ideal glassy state (if the liquid has the infinite time needed to equilibrate). But $T_K$ is identified by extrapolation, and one may well ask how reliable the extrapolation is. This question arises, in particular, if one accepts that $T_g$ is close to a genuine phase transition as predicted by the AG model. It seems quite possible that the liquid entropy may ``bend over'' and stay above the crystalline entropy right down to zero temperature \cite{guj81,sti01,eck08}. This would imply $T_K=0$.

Suppose that we nevertheless accept that data conform to Eq. (\ref{tkeqto}) -- the AG model's intriguing linking of a purely dynamic temperature ($T_0$) to a purely thermodynamic one ($T_K$). Very recently the VFT equation's predicted divergence was questioned in a paper that compiled accurate data for the dielectric relaxation time's temperature dependence for 42 organic liquids \cite{hec08}. The conclusion was that, while the VFT equation does work well as a simple representation of data, there is no evidence for any dynamic divergence; in other words, there is little evidence that $T_0$ exists \cite{sim01}.

Suppose that we nevertheless accept both the extrapolation usually carried out in order to identify $T_K$ and the existence of the VFT $T_0$. Then a simple experimental test of the entropy model is the prediction Eq. (\ref{tkeqto}). Numerous papers published the last 30 years have reported confirmation of Eq. (\ref{tkeqto}); indeed this appears to be one of the strongest experimental argument for the entropy model. In 2003, however, Tanaka compiled a large amount of data and concluded that Eq. (\ref{tkeqto}) is disobeyed \cite{tan03}.

\section{Concluding remarks}

The classical Adam-Gibbs scenario presents several challenges. Thus if entropy is the central variable controlling the relaxation time's temperature dependence, it seems that more advanced approaches are needed. In our opinion it is more likely, however, that entropy will not maintain its central role in theories of viscous liquid dynamics. Even if one limits the search to phenomenological models, there are viable alternatives like the elastic models that date back to the 1940's \cite{dyr06}. According to the shoving model \cite{dyr96}, one of the elastic models, the activation energy is proportional to the instantaneous shear modulus $G_\infty$. This quantity is quite temperature dependent in viscous liquids, in fact enough to explain the non-Arrhenius behavior. Since $G_\infty$ cannot diverge, there is no underlying phase transition, so the elastic model scenario differs qualitatively from that of the entropy model. It will be interesting to see which of the two ideas prevails -- if any.

\end{document}